\documentclass[epj]{svjour}
\usepackage{graphics}
\usepackage{psfig}
\begin{document}
\title{Phenomenology of the $SU(3)_c\otimes SU(3)_L\otimes U(1)_X$ model 
with right-handed neutrinos}
\author{Diego A. Guti\'errez\inst{1} \and William A. Ponce\inst{1} \and Luis A. S\'anchez\inst{2}}                     
\institute{Instituto de F\'\i sica, Universidad de Antioquia,
A.A. 1226, Medell\'\i n, Colombia \and Escuela de F\'\i sica, Universidad Nacional de Colombia,
A.A. 3840, Medell\'\i n, Colombia}
\date{Received: date / Revised version: date}
%
\abstract{
A phenomenological analysis of the three-family model based on the local gauge group
$SU(3)_c\otimes SU(3)_L\otimes U(1)_X$ with right-handed neutrinos, is
carried out. Instead of using the minimal scalar sector able to break the
symmetry in a proper way, we introduce an alternative set of four Higgs scalar triplets, which combined with an anomaly-free discrete symmetry, produces a quark mass spectrum without hierarchies in the Yukawa coupling constants. We also embed the structure into a simple gauge group and show some conditions to achieve a low energy gauge coupling unification, avoiding possible conflict with proton decay bounds. By using experimental results from the CERN-LEP, SLAC linear collider, and atomic parity violation data, we update constraints on several parameters of the model.
 } 
\maketitle
\section{Introduction}
\label{intro}
Two intriguing puzzles completely unanswered in modern particle physics are the 
number of fermion families in nature, and the pattern of masses and
mixing angles in the fermion sector. 
One interesting attempt to answer to the question of family
replication is provided by the 3-3-1 extension of the local gauge symmetry 
$SU(3)_c\otimes SU(2)_L\otimes U(1)_Y$ of the Standard Model (SM)
of the strong and electroweak interactions~\cite{sm}. This extension, based on the
local gauge group $SU(3)_c\otimes SU(3)_L\otimes U(1)_X$, has among its
best features that several models can be constructed so that anomaly
cancellation is achieved by an interplay between the families, all of them
under the condition $N_f=N_c=3$, where $N_f$ is the number of families and
$N_c$ is the number of colors of $SU(3)_c$ (three-family models)
\cite{pf}.

Two 3-3-1 three-family models have been extensively studied over the last
decade \cite{pf,vl}. In one of them the three known left-handed lepton
components for each family are associated to three $SU(3)_L$ triplets as
$(\nu_l,l^-,l^+)_L$, where $l^+_L$ is related to the right-handed isospin
singlet of the charged lepton $l^-_L$ in the SM \cite{pf}. In the other
model the three $SU(3)_L$ lepton triplets are of the form $(\nu_l, l^-,
\nu_l^c)_L$, where $\nu_l^c$ is related to the right-handed component of
the neutrino field $\nu_l$ (a model with right-handed neutrinos)  
\cite{vl}. In the first model anomaly cancellation implies quarks with
exotic electric charges $-4/3$ and $5/3$, while in the second one the extra
particles have only ordinary electric charges.

Our aim in this paper is to do a phenomenological analysis of the 3-3-1
model in the version that includes right-handed neutrinos, including a detailed study of the fermion mass spectrum, with emphasis in the quark sector. Previous works~\cite{vl} just present the Yukawa Lagrangians without looking for constraints able to produce a consistent quark mass spectrum. 
It will be shown that a convenient set of four Higgs scalars, combined with an appropriate anomaly-free discrete $Z_2$ symmetry, produces an appealing quark mass spectrum without strong hierarchies for the Yukawa couplings. 
Besides, we are going to study the embedding and unification of this gauge structure into $SU(6)$, which is an appropriate unification gauge group. Finally we will set updated constraints on several parameters of the model. 

The problem of the lepton masses in the context of 3-3-1 three-family models has been studied, for example, in Refs.~\cite{mpp1,mpp2}, and we already know, from the analysis presented in Refs.~\cite{mpp2,kita1,kita2},
that models based on the 3-3-1 local gauge structure are suitable to
describe some neutrino properties, because they include in a natural way
most of the ingredients needed to explain the masses and mixing in the
neutrino sector. In particular, Ref.~\cite{kita1} addresses this issue for the model studied here.

This paper is organized as follows: in Sect.~\ref{sec:1} we review
the model, introduce the new scalar sector, embed the structure into a 
covering group and calculate the charged and neutral electroweak currents; in Sect.~\ref{sec:2} we study the charged fermion mass spectrum; in Sect.~\ref{sec:3} we do the renormalization group equation analysis and show the conditions for the gauge coupling unification; in Sect.~\ref{sec:4} we fix the new bounds on the mixing angle between the two flavor diagonal neutral currents present in the model, and discuss the constraints coming from violation of unitarity of the Cabbibo-Kobayashi-Maskawa (CKM) quark-mixing matrix and from Flavour Changing Neutral Currents (FCNC). Finally, in the last Section, we present our conclusions.

\section{The Model}
\label{sec:1}
The model we are about to study here was sketched for the first time in the literature in the first reference in \cite{vl}, with some phenomenology presented in the other four papers in the same reference. Some of the formulas quoted in the following Sections are taken from those references and from Ref.~\cite{pgs}; corrections to some minor printing mistakes in the original papers are included.

\subsection{The gauge group} 
As it was stated above, the model we are interested in, is based on the
local gauge group $SU(3)_c\otimes SU(3)_L\otimes U(1)_X$ which has 17
gauge bosons: one gauge field $B^\mu$ associated with $U(1)_X$, the 8
gluon fields $G^\mu$ associated with $SU(3)_c$ which remain massless after
breaking the symmetry, and another 8 gauge fields associated with
$SU(3)_L$ and that we write for convenience as \cite{pgs}

\[{1\over 2}\lambda_\alpha A^\mu_\alpha={1\over \sqrt{2}}\left(
\begin{array}{ccc}D^\mu_1 & W^{+\mu} & K^{+\mu} \\ W^{-\mu} & D^\mu_2 &
K^{0\mu} \\ K^{-\mu} & \bar{K}^{0\mu} & D^\mu_3 \end{array}\right), \]
where $D^\mu_1=A_3^\mu/\sqrt{2}+A_8^\mu/\sqrt{6},\;
D^\mu_2=-A_3^\mu/\sqrt{2}+A_8^\mu/\sqrt{6}$, and
$D^\mu_3=-2A_8^\mu/\sqrt{6}$. $\lambda_i, \; i=1,2,...,8$, are the eight
Gell-Mann matrices normalized as $Tr(\lambda_i\lambda_j)  
=2\delta_{ij}$.

The charge operator associated with the unbroken gauge symmetry $U(1)_Q$ 
is given by
\begin{equation}
Q=\frac{\lambda_{3L}}{2}+\frac{\lambda_{8L}}{2\sqrt{3}}+XI_3,
\end{equation}
where $I_3=Diag.(1,1,1)$ is the diagonal $3\times 3$ unit matrix, and the 
$X$ values are related to the $U(1)_X$ hypercharge which are fixed by 
anomaly cancellation. 
The sine square of the electroweak mixing angle is given by
\begin{equation} \label{sinteta}
S_W^2=3g_1^2/(3g_3^2+4g_1^2),
\end{equation}
where $g_1$ and $g_3$ are the gauge coupling constants of $U(1)_X$ and $SU(3)_L$ respectively, and the photon field is 
given by \cite{vl,pgs}
\begin{equation}\label{foton}
A_0^\mu=S_WA_3^\mu+C_W\left[\frac{T_W}{\sqrt{3}}A_8^\mu + 
\sqrt{(1-T_W^2/3)}B^\mu\right],
\end{equation}
where $C_W$ and $T_W$ are the cosine and tangent of the electroweak mixing 
angle, respectively. 

There are two weak neutral currents in the model 
associated with the two flavor diagonal neutral weak gauge bosons 
\begin{eqnarray}\nonumber \label{zzs}
Z_0^\mu&=&C_WA_3^\mu-S_W\left[\frac{T_W}{\sqrt{3}}A_8^\mu + 
\sqrt{(1-T_W^2/3)}B^\mu\right], \\ \label{zetas}
Z_0^{\prime\mu}&=&-\sqrt{(1-T_W^2/3)}A_8^\mu+\frac{T_W}{\sqrt{3}}B^\mu,
\end{eqnarray}
and one current associated with the flavor non diagonal electrically neutral gauge boson $K^{0\mu}$, which carries a kind of weak 
V-isospin charge. In the former expressions $Z^\mu_0$ coincides with the weak neutral current of the SM \cite{vl,pgs}. Using  
Eqs.~(\ref{foton}) and (\ref{zetas}) we may realize that the gauge boson 
$Y^\mu$ associated with the abelian hypercharge in the $SU(3)_c\otimes SU(2)_L\otimes U(1)_Y$  SM gauge group is  
\begin{equation} \label{hyper}
Y^\mu=\frac{T_W}{\sqrt{3}}A_8^\mu + 
\sqrt{(1-T_W^2/3)}B^\mu.
\end{equation}

\subsection{The spin 1/2 particle content} 
The quark content for the three families in this model (known in the literature as the 3-3-1 model with right-handed neutrinos) is the following:  
$Q^i_{L}=(u^i,d^i,D^i)_L\sim(3,3,0)$, $i=1,2$ for two families, where
$D^i_L$ are two extra quarks of electric charge $-1/3$ (the numbers
inside the parentheses stand for the $[SU(3)_c,SU(3)_L,U(1)_X]$ quantum
numbers in that order); $Q^3_{L}=(d^3,u^3,U)_L\sim (3,3^*,1/3)$, where
$U_L$ is an extra quark of electric charge 2/3. The right handed quarks
are $u^{ac}_{L}\sim (3^*,1,-2/3)$, $d^{ac}_{L}\sim (3^*,1,1/3)$ with
$a=1,2,3$ a family index, $D^{ic}_{L}\sim (3^*,1,1/3)$, $i=1,2$, and
$U^c_L\sim (3^*,1,-2/3)$.

The lepton content is given by the three $SU(3)_L$ anti-triplets $L_{lL} =
(l^-,\nu_l^0,\nu_l^{0c})_L\sim (1,3^*,-1/3)$, for $l=e,\mu,\tau$ 
a leptonic family index, and the three singlets $l^+_{L}\sim(1,1,1)$, 
where $\nu_l^0$ is the neutrino field associated with the lepton $l^-$, and $\nu_l^{0c}$ plays the role of the right-handed neutrino field associated to the same flavor. Notice that this model does not contain exotic charged leptons, and universality for the known leptons in the three families is present at tree level in the weak basis.
 
With these quantum numbers it is just a matter of counting to check
that the model is free of the following gauge anomalies \cite{pgs}:  
$[SU(3)_c]^3$; (as in the SM, $SU(3)_c$ is vectorlike); $[SU(3)_L]^3$ (six
triplets and six anti-triplets), $[SU(3)_c]^2U(1)_X$; $[SU(3)_L]^2U(1)_X$; $[grav]^2U(1)_X$ and $[U(1)_X]^3$, where $[grav]^2U(1)_X$ stands for
the gravitational anomaly as described in Ref.~\cite{del}.

\subsection{The new scalar sector}
Instead of using the set of three triplets of Higgs scalars introduced in the original papers \cite{vl}, or the most economical set of two triplets introduced in Ref.~\cite{pgs} (none of them able to produce a realistic mass spectrum), we propose here to work with the following set of four Higgs scalar fields, and Vacuum Expectation Values (VEV):
\begin{eqnarray}\label{higgsses} \nonumber
\langle\phi_1^T\rangle &=&\langle(\phi^+_1, \phi^0_1,\phi^{'0}_1)\rangle = 
\langle(0,0,v_1)\rangle \sim (1,3,1/3) \\ \nonumber
\langle\phi_2^T\rangle &=&\langle(\phi^+_2, \phi^0_2,\phi^{'0}_2)\rangle = 
\langle(0,v_2,0)\rangle \sim (1,3,1/3) \\ \nonumber
\langle\phi_3^T\rangle &=&\langle(\phi^0_3, \phi^-_3,\phi^{'-}_3)\rangle = 
\langle(v_3,0,0)\rangle \sim (1,3,-2/3) \\ \nonumber
\langle\phi_4^T\rangle &=&\langle(\phi^+_4, \phi^0_4,\phi^{'0}_4)\rangle = 
\langle(0,0,V)\rangle \sim (1,3,1/3), \\
\end{eqnarray}
with the hierarchy $v_1\sim v_2\sim v_3\sim 10^2$ GeV $<< V$. 
The analysis shows that this set of VEV breaks the 
$SU(3)_c\otimes SU(3)_L\otimes U(1)_X$ symmetry in two steps 
following the scheme
\begin{eqnarray*}
SU(3)_c\otimes SU(3)_L\otimes U(1)_X &\stackrel{(V+v_1)}{\longrightarrow}& \\
SU(3)_c\otimes SU(2)_L\otimes U(1)_Y&\stackrel{(v_2+v_3)}{\longrightarrow}& SU(3)_c\otimes U(1)_Q,
\end{eqnarray*} 
which in turn allows for the matching conditions $g_2=g_3$ and 
\begin{equation}\label{mc}
\frac{1}{g^{2}_Y}=\frac{1}{g_1^2}+\frac{1}{3g_2^2},
\end{equation}
where $g_2$ and $g_Y$ are the gauge coupling constants of 
the $SU(2)_L$ and $U(1)_Y$ gauge groups in the SM, respectively. 

We will see in the next Sections that this scalar structure 
properly breaks the symmetry, provides with masses for the gauge 
bosons and, combined with a discrete symmetry, it is enough to produce a 
consistent mass spectrum for the up and down quark sectors (a realistic mass spectrum in the lepton sector requires new ingredients as for example $SU(3)_L$ leptoquark scalar triplets and/or sextuplets, as we will briefly mention ahead).

\subsection{$SU(6)\supset SU(5)$ as a covering group}
The Lie algebra of $SU(3)\otimes SU(3)\otimes U(1)$ is a maximal  
subalgebra of the simple algebra of $SU(6)$. The five fundamental 
irreducible representations (irreps) of $SU(6)$ are: 
$\{6\},\{6^*\},\{15\},\{15^*\}$ and the \{20\} which is real. The branching rules for 
these fundamental irreps into 
$SU(3)_c\otimes SU(3)_L\otimes U(1)_X$ are \cite{slansky}:
\begin{eqnarray}\label{branching}\nonumber
\{6\} &\rightarrow &(3,1,-1/3)\oplus (1,3,1/3), \\ \nonumber
\{15\} &\rightarrow & (3^*,1,-2/3)\oplus (1,3^*,2/3) \oplus (3,3,0), \\ \nonumber
\{20\} &\rightarrow & (1,1,1)\oplus (1,1,-1)\oplus (3,3^*,1/3)\oplus (3^*,3,-1/3),
\end{eqnarray}
where we have normalized the $U(1)_X$ hypercharge according to our convenience.

From these branching rules and from the fermion structure presented above, it is clear that all the particles in the 3-3-1 model with right-handed neutrinos can be included in the following $SU(6)$ reducible representation
\begin{equation}\label{reducible}
5\{6^*\}+3\{20\}+4\{15\},
\end{equation}
which includes new exotic particles, as for example 
\begin{eqnarray*}
(N^0,E^+,E^{\prime +})_L&\sim&(1,3^*,2/3)\subset \{15\}, \\
E^-_L&\sim&(1,1,-1)\subset \{20\}, \\
(D^{\prime c},U^{\prime c},U^{\prime\prime c})_L&\sim&(3^*,3,-1/3)\subset 
\{20\}. \end{eqnarray*}

The analysis reveals that the reducible representation in (\ref{reducible}) is anomalous. The simplest $SU(6)$ reducible representation which is free of anomalies and includes the fields in  
(\ref{reducible}) is given by \cite{slansky} 
\begin{equation}\label{rred}
8\{6^*\}+3\{20\}+4\{15\},
\end{equation}
which also includes the following new exotic particles (all of them with 
ordinary electric charges): four families of 3-3-1 up- and down-type
quarks, four more exotic down-type quarks, plus eight families of 3-3-1 
lepton triplets, among a good deal of other particles.

It is clear from the following decomposition of irrep $\{6^*\}$ of $SU(6)$ into $SU(5)\otimes U(1)$ 
\begin{equation}
\{6^*\}=\{d^c,-N_E^0,E^-,N^{0c}_E\}_L\longrightarrow \{d^c,-N^0_E,E^-\}_L\oplus N^{0c}_{EL}, 
\end{equation}
that for $N^0_{EL}=\nu_{eL}$ and $E^-_L=e^-_L$, we obtain the known $SU(5)$ model of Georgi and Glashow \cite{gg}; so, in some sense, this model is an extension of one of the first Grand Unified Theories (GUT) presented in the literature.

\subsection{The gauge boson sector}
After breaking the symmetry with $\langle\phi_i\rangle,\; i=1,\dots ,4$, and using the covariant derivative for triplets $D^\mu=\partial^\mu
-ig_3\lambda_{\alpha L}A^\mu_\alpha/2-ig_1XB_\mu I_3$, we get the 
following mass terms in the gauge boson sector.

\subsubsection{Spectrum in the charged gauge boson sector}
A straightforward calculation shows that the charged gauge bosons $K^\pm_\mu$ and $W^\pm_\mu$ do not mix with each other and get the following masses: $M^2_{K^\pm}=g_3^2(V^2+v_1^2+v_3^2)/2$ and $M_W^2=g_3^2(v_2^2+v_3^2)/2$, which for $g_3=g_2$ and using the experimental value $M_{W}=80.423 \pm 0.039$ GeV \cite{pdb} implies $\sqrt{v_2^2+v_3^2}\simeq 175$ GeV. In the same way $K^{0\mu}$ (and its antiparticle $\bar{K}^{0\mu}$) does not mix with the other two electrically neutral gauge bosons and gets a bare mass $M_{K^0}^2=g_3^2(V^2+v_1^2+v_2^2)/2\approx M^2_{K^\pm}$. Notice that $v_1$ does not contribute to the $W^\pm$ mass because it is associated with an $SU(2)_L$ singlet Higgs scalar.

\subsubsection{Spectrum in the neutral gauge boson sector}
The algebra now shows that in this sector the photon field $A_0^\mu$ in Eq.~(\ref{foton}) decouples from $Z_0^\mu$ and $Z_0^{\prime\mu}$ and remains massless. Then, in the basis $(Z_0^\mu,Z_0^{\prime\mu})$, we obtain the following $2\times 2$ mass matrix 
\begin{equation} \label{gauges}
\frac{\eta^2g_3^2}{4C_W^2}\left( \begin{array}{cc}
\frac{v_2^2+v_3^2}{\eta^2}& 
\frac{v_2^2C_{2W}-v_3^2}{\eta} \\
\frac{v_2^2C_{2W}-v_3^2}{\eta} & 
v_2^2C^2_{2W}+ v_3^2+4(V^2+v_1^2)C_W^4
\end{array}\right),
\end{equation}
where $C_{2W}=C_W^2-S_W^2$ and $\eta^{-2}=(3-4S_W^2)$. This matrix provides with a mixing between $Z^\mu_0$ and $Z^{\prime\mu}_0$ of the form 
\begin{eqnarray} \label{tan} \nonumber
\tan(2\theta)&= \frac{2\sqrt{(3 - 4S^2_W)}(v_2^2C_{2W}-v_3^2)}
{4C_W^4(V^2+v_1^2) - 2v_3^2C_{2W}-v_2^2(3-4S_W^2-C_{2W}^2)}\\ 
&\stackrel{V\rightarrow\infty}{\longrightarrow}0 .
\end{eqnarray}
The physical fields are then 

\begin{eqnarray}\nonumber
Z_1^\mu&=&Z^\mu_0 
\cos\theta-Z^{\prime\mu}_0\sin\theta \; ,\\ \nonumber
Z_2^\mu&=&Z^\mu_0 \sin\theta+Z'^\mu_0 \cos\theta .
\end{eqnarray} 
An updated bound on the mixing angle $\theta$ is going to be calculated in 
Sect.~\ref{sec:4} using experimental results.

\subsection{Currents}
\subsubsection{Charged currents}
The Hamiltonian for the currents, charged under the generators of the 
$SU(3)_L$ group, is  
$H^{CC}=g_3(W^+_\mu J^\mu_{W^+}+K^+_\mu J_{K^+}^\mu+K^0_\mu 
J_{K^0}^\mu)/\sqrt{2}+h.c.$, with

\begin{eqnarray}\nonumber
J_{W^+}^\mu&=& (\sum_{i=1}^2\bar{u}^i_{L}\gamma^\mu d^i_{L})
-\bar{u}^3_{L}\gamma^\mu d^3_{L} - \sum_{l=e,\mu,\tau}
\bar{\nu}_{l L}\gamma^\mu l^-_{L},\\ \nonumber 
J^\mu_{K^+}&=&(\sum_{i=1}^2\bar{u}^i_{L}\gamma^\mu 
D^i_{L})-\bar{U}_{L}\gamma^\mu 
d^3_{L} - \sum_{l=e,\mu,\tau}
\bar{\nu}^{0c}_{l L}\gamma^\mu l^-_{L}, \\ \nonumber 
J^\mu_{K^0}&=&(\sum_{i=1}^2\bar{d}^i_{L}\gamma^\mu 
D^i_{L})-\bar{U}_{L}\gamma^\mu u^3_{L} - \sum_{l=e,\mu,\tau}
\bar{\nu}^{0c}_{l L}\gamma^\mu \nu_{l L}, \nonumber
\end{eqnarray} 
where $K^0_\mu$ is an electrically neutral gauge boson, but it carries a kind of weak V-isospin charge, besides, it is flavor non diagonal.

\subsubsection{Neutral currents}
The neutral currents $J_\mu(EM)$, $J_\mu(Z)$ and $J_\mu(Z^\prime)$,
associated with the Hamiltonian
\begin{equation} 
H^0 = eA^\mu J_\mu(EM)+(g_3/{C_W})Z^\mu
J_\mu(Z)+ (g_1/\sqrt{3})Z^{\prime\mu}J_\mu(Z^\prime), 
\end{equation}
are~\cite{vl}

\begin{eqnarray}\nonumber
J_\mu(EM)&=&{2\over 3}\left[\sum_{a=1}^3\bar{u}_a\gamma_\mu u_a +
\bar{U}\gamma_\mu U \right] \\ \nonumber
& &- {1\over 3}\left[\sum_{a=1}^3\bar{d}^a\gamma_\mu d^a+ 
\sum_{i=1}^2\bar{D}^i\gamma_\mu D^i \right]  \\ \nonumber
& &- \sum_{l=e,\mu,\tau}\bar{l}^-\gamma_\mu l^- \\ \nonumber
& =& \sum_f q_f\bar{f}\gamma_\mu f,\\* \nonumber
J_\mu(Z)&=&J_{\mu,L}(Z)-S^2_WJ_\mu(EM),\\ \nonumber
J_\mu(Z^\prime)&=& - J_{\mu,L}(Z^\prime)+T_WJ_\mu(EM), 
\end{eqnarray}
where $e=g_3S_W=g_1C_W\sqrt{(1-T_W^2/3)}>0$ is the electric charge, 
$q_f$ is the electric charge of the fermion $f$ in units of $e$, and 
$J_\mu(EM)$ is the electromagnetic current. 

The left-handed currents are
\begin{eqnarray} \nonumber
J_{\mu,L}(Z)&=&{1\over 2}[\sum_{a=1}^3(\bar{u}^a_{L}\gamma_\mu u^a_{L}
-\bar{d}^a_{L}\gamma_\mu d^a_{L})\\ \nonumber
& &+ \sum_{l=e,\mu,\tau}(\bar{\nu}_{l L}\gamma_\mu \nu_{l L} 
-\bar{l}^-_{L}\gamma_\mu l^-_{L})] \\ \label{jotaz}
&=&\sum_F \bar{F}_LT_{3f}\gamma_\mu F_L ,
\end{eqnarray}
\begin{eqnarray}\nonumber
J_{\mu,L}(Z^\prime)&=& 
S^{-1}_{2W}[\bar{u}_{1L}\gamma_\mu u_{1L}+\bar{u}_{2L}\gamma_\mu u_{2L}
\\ \nonumber
& &-\bar{d}_{3L}\gamma_\mu d_{3L}
-\sum_l(\bar{l}^-_{L}\gamma_\mu l^-_{L})] \\ \nonumber
& &+T^{-1}_{2W}[\bar{d}_{1L}\gamma_\mu d_{1L}+\bar{d}_{2L}\gamma_\mu 
d_{2L}\\ \nonumber
& &-\bar{u}_{3L}\gamma_\mu u_{3L}-\sum_l(\bar{\nu}_{lL}\gamma_\mu \nu_{lL})] \\  \nonumber
& &+T^{-1}_{W}[\bar{D}_{1L}\gamma_\mu D_{1L}+\bar{D}_{2L}\gamma_\mu 
D_{2L}\\ \nonumber
& &-\bar{U}_{L}\gamma_\mu U_{L}-\sum_l(\bar{\nu}^{0c}_{lL}\gamma_\mu \nu^{0c}_{lL})] \\ \label{jzprima}
&=&\sum_F\bar{F}_L T^\prime_{3f}\gamma_\mu F_L,
\end{eqnarray}
where $S_{2W}=2S_WC_W$, $T_{2W}=S_{2W}/C_{2W}$, $T_{3f}$ 
= $Dg(1/2,-1/2,0)$ is the third component of the weak isospin, $T^\prime_{3f}$ = $Dg(S^{-1}_{2W},T^{-1}_{2W},-T_W^{-1})$ is a convenient $3\times 3$ diagonal matrix, acting both of them on the representation 3 
of $SU(3)_L$ (the negative value when acting on the representation $3^*$, which is also true for the matrix $T_{3f}$) and $F$ is a generic
symbol for the representations 3 and $3^*$ of $SU(3)_L$.  Notice that
$J_\mu(Z)$ is the neutral current of the SM (with the extra fields 
included in $J_\mu (EM$)). This allows us to identify $Z_\mu$ as the 
neutral gauge boson of the SM, which is consistent with Eqs.~(\ref{zzs}) 
and (\ref{hyper}).

The couplings of the flavor diagonal mass eigenstates $Z_1^\mu$ and 
$Z_2^\mu$ are given by
\begin{eqnarray} \nonumber
H^{NC}&=&\frac{g_3}{2C_W}\sum_{i=1}^2Z_i^\mu\sum_f\{\bar{f}\gamma_\mu
[a_{iL}(f)(1-\gamma_5)\\ \nonumber & & 
+a_{iR}(f)(1+\gamma_5)]f\} \\ \nonumber
      &=&\frac{g_3}{2C_W}\sum_{i=1}^2Z_i^\mu\sum_f\{\bar{f}\gamma_\mu
      [g(f)_{iV}-g(f)_{iA}\gamma_5]f\},
\end{eqnarray}
with
\begin{eqnarray} \nonumber
a_{1L}(f)&=&\cos\theta (T_{3f}-q_fS^2_W)\\ \nonumber & &
+\Theta\sin\theta 
(T^\prime_{3f}-q_fT_W), \\ \nonumber
a_{1R}(f)&=&-q_f\left(\cos\theta S_W^2
+\Theta\sin\theta T_W\right),\\ \nonumber
a_{2L}(f)&=&\sin\theta (T_{3f}-q_fS^2_W)\\ \nonumber & &
-\Theta\cos\theta  
(T^\prime_{3f}-q_fT_W),\\ 
\label{a}
a_{2R}(f)&=&-q_f\left(\sin\theta S^2_W-\Theta\cos\theta T_W\right),
\end{eqnarray}
where $\Theta = S_WC_W/\sqrt{(3-4S_W^2)}$.
From these coefficients we can read 
\begin{eqnarray} \nonumber
g(f)_{1V}&=&\cos\theta (T_{3f}-2q_fS^2_W)\\ \nonumber & &
+\Theta\sin\theta (T^\prime_{3f}-2q_fT_W), \\ \nonumber
g(f)_{2V}&=&\sin\theta (T_{3f}-2q_fS^2_W)\\ \nonumber & &
-\Theta\cos\theta  
(T^\prime_{3f}-2q_fT_W),\\ 
g(f)_{1A}&=&\cos\theta  
T_{3f}+\Theta\sin\theta T^\prime_{3f}, \\ \nonumber \label{g}
g(f)_{2A}&=&\sin\theta T_{3f}-\Theta\cos\theta T^\prime_{3f}.
\end{eqnarray}
The values of $g_{iV}$ and $g_{iA}$, with $i=1,2$, are listed in Tables~\ref{tab:1} and \ref{tab:2}~\cite{vl}.

\begin{table*}
\caption{The $Z_1^\mu\longrightarrow \bar{f}f$ couplings.}
\label{tab:1}
\begin{tabular}{lcc}
\hline\noalign{\smallskip}
$f$ & $g(f)_{1V}$ & $g(f)_{1A}$ \\ 
\noalign{\smallskip}\hline\noalign{\smallskip}
$u^{1,2}$& 
$({1\over 2}-{4S_W^2 \over 3})\cos\theta
+\Theta(s_{2W}^{-1}-{4T_W \over 3})\sin\theta$
& ${1\over 2}\cos\theta + \Theta S_{2W}^{-1}\sin\theta$  \\ 
$u^{3}$&$({1\over 2}-{4S_W^2 \over 3})\cos\theta -  
\Theta (T_{2W}^{-1}+{4T_W\over 3})\sin\theta$ & 
${1\over 2}\cos\theta - \Theta T_{2W}^{-1}\sin\theta$\\
$d^{1,2}$ & $(-{1\over 2}+{2S_W^2\over 3})\cos\theta 
+\Theta(T_{2W}^{-1}+{2T_W\over 3})\sin\theta$ 
& $-{1\over 2}\cos\theta +\Theta T_{2W}^{-1}\sin\theta$ \\
$d^{3}$ & $(-{1\over 2}+{2S_W^2\over 3})\cos\theta
-\Theta(S_{2W}^{-1}-{2T_W\over 3})\sin\theta$ & 
$-{1\over 2}\cos\theta - \Theta S_{2W}^{-1}\sin\theta$\\
$U$ & $-{4S_W^2\over 3}\cos\theta-\Theta(T_W^{-1}+
{4T_W\over 3})\sin\theta $ & $\Theta T_W^{-1}\sin\theta $ \\
$D^{1,2}$ & ${2S_W^2\over 3}\cos\theta
+\Theta (T_W^{-1}+{2T_W\over 3})\sin\theta $ &
$-\Theta T_W^{-1}\sin\theta $ \\
$e,\mu,\tau$& $(-{1\over 2}+2S_W^2)\cos\theta -
\Theta(S_{2W}^{-1}-2T_W)\sin\theta $ 
& $ -{1\over 2}\cos\theta -\Theta S_{2W}^{-1}\sin\theta $\\
$\nu_e,\nu_\mu,\nu_\tau$& ${1\over 2}\cos\theta -\Theta 
T_{2W}^{-1}\sin\theta$ 
& ${1\over 2}\cos\theta -\Theta T_{2W}^{-1}\sin\theta$ \\
$\nu^{0c}_e,\nu^{0c}_\mu,\nu^{0c}_\tau$ & $-\Theta T_W^{-1}\sin\theta $ &
                $-\Theta T_W^{-1}\sin\theta $ \\ 
\noalign{\smallskip}\hline
\end{tabular}
\end{table*}

\begin{table*}
\caption{The $Z_2^\mu\longrightarrow \bar{f}f$ couplings.}
\label{tab:2}
\begin{tabular}{lcc}
\hline\noalign{\smallskip}
$f$ & $g(f)_{2V}$ & $g(f)_{2A}$ \\ 
\noalign{\smallskip}\hline\noalign{\smallskip}
$u^{1,2}$& $({1\over 2}-{4S_W^2 \over 3})\sin\theta -\Theta (S_{2W}^{-1}-
{4T_W\over 3})\cos\theta $
& ${1\over 2}\sin\theta -\Theta S_{2W}^{-1}\cos\theta$ \\ 
$u^{3}$&$({1\over 2}-{4S_W^2 \over 3})\sin\theta +\Theta (T_{2W}^{-1}+ {4T_W\over 
3})\cos\theta $
& ${1\over 2}\sin\theta+\Theta T_{2W}^{-1}\cos\theta $\\
$d^{1,2}$ & $(-{1\over 2}+{2S_W^2\over 3})\sin\theta -
\Theta(T_{2W}^{-1}+{2T_W\over 3})\cos\theta$
& $-{1\over 2}\sin\theta -\Theta T_{2W}^{-1}\cos\theta $ \\
$d^{3}$ & $(-{1\over 2}+{2S_W^2\over 3})\sin\theta +
\Theta (S_{2W}^{-1}-{2T_W\over 3})\cos\theta$ & 
$-{1\over 2}\sin\theta +\Theta S_{2W}^{-1}\cos\theta$\\
$U$ & $-{4S_W^2\over 3}\sin\theta + 
\Theta(T_W^{-1}+{4T_W\over 3})\cos\theta$ & 
$\Theta T_W^{-1}\cos\theta$\\
$D^{1,2}$ & ${2S_W^2\over 3}\sin\theta 
-\Theta(T_W^{-1}+{2T_W\over 3})\cos\theta$ &
$-\Theta T_W^{-1}\cos\theta$ \\
$e,\mu,\tau$& $(-{1\over 2}+2S_W^2)\sin\theta +\Theta(S_{2W}^{-1}-
{2T_W\over 3})\cos\theta$ & 
$-{1\over 2}\sin\theta +\Theta S_{2W}^{-1}\cos\theta$\\
$\nu_e,\nu_\mu,\nu_\tau$ & ${1\over 2}\sin\theta +\Theta T_{2W}^{-1}\cos\theta$ & 
${1\over 2}\sin\theta +\Theta T_{2W}^{-1}\cos\theta$ \\
$\nu^{0c}_e,\nu^{0c}_\mu,\nu^{0c}_\tau$ & $\Theta T_W^{-1}\cos\theta$ &
$\Theta T_W^{-1}\cos\theta$ \\
\noalign{\smallskip}\hline
\end{tabular}
\end{table*}

As we can see, in the limit $\theta =0$ the couplings of
$Z_1^\mu$ to the ordinary leptons and quarks are the same as in the SM;
due to this property we can test the new physics beyond the SM predicted 
by this particular model.

\section{Fermion Masses}
\label{sec:2}
The Higgs scalars introduced in Sect.~\ref{sec:1} break the symmetry in 
an appropriate way and, at the same time, produce mass terms for the fermion fields via Yukawa interactions. 

In order to restrict the number of Yukawa couplings, and
produce a realistic mass spectrum, we introduce an anomaly-free discrete $Z_2$
symmetry~\cite{ross} with the following assignments of charges:
\begin{eqnarray}\label{z2} \nonumber
Z_2(Q^a_L,\phi_2,\phi_3,\phi_4,u^{ic}_L,d^{ac}_L)&=&1\\ 
Z_2(\phi_1, u^{3c}_L,U^c_L,D^{ic}_L, L_{lL}, l_{L}^+)&=&0,
\end{eqnarray}
where $a=1,2,3, \; i=1,2$ and $l=e,\mu,\tau$ are family indexes as above.

\subsection{The up quark sector}
The most general invariant Yukawa Lagrangian for the up quark sector is 
given by

\begin{eqnarray}\label{mup}
{\cal L}^u_Y&=&
\sum_{\alpha=1,2,4}Q_L^3\phi_\alpha C(h^U_\alpha
U_L^c+\sum_{a=1}^3h_{a\alpha}^uu_L^{ac})\\ \nonumber
&+& \sum_{i=1}^2Q^i_L\phi_3^*
C(\sum_{a=1}^3h^u_{ia}u_L^{ac}+h_i^{\prime U}U_L^c)
+ h.c.,
\end{eqnarray}
where the $h'$s are Yukawa coupling constants and $C$ is the charge conjugation operator. 

Then, in the basis $(u^1,u^2,u^3,U)$, and using the $Z_2$ symmetry,  we get from Eqs.~(\ref{z2},\ref{mup}) the following tree-level up quark mass matrix 

\begin{equation} \label{matrixU}
M_u=\left(\begin{array}{cccc}
0 & 0 & 0 & h^u_{11}v_1\\
0 & 0 & 0 & h^u_{21}v_1\\
h^u_{13}v_3 & h^u_{23}v_3 & h^u_{32}v_2 & h_{34}^uV\\
h_1^{\prime U}v_3 & h^{\prime U}_2v_3 & h^U_2v_2 & h^U_4V \\
\end{array}\right), \end{equation}
which is a rank one see-saw type mass matrix. As a matter of fact, analytical and numerical analysis of this matrix shows that 
$M_u^\dagger M_u$ has one eigenvalue equal to zero related to the
eigenvector $[(h_{32}^uh_{2}^{\prime U}-h_{23}^uh_{2}^U),
(h_{13}^uh_{2}^U-h_{32}^uh_{1}^{\prime U}), (h_{23}^uh_{1}^{\prime
U}-h_{13}^uh_{2}^{\prime U}),0]$, that we may identify with the up
quark $u$ in the first family, which remains massless at the tree-level.

In what follows, and without loss of generality, we are going to
impose the condition $v_1=v_2=v_3\equiv v << V$, with the value
for $v$ fixed by the mass of the charged weak gauge boson
$M_{W^\pm}^2=g_3^2(v_2^2+v_3^2)/2=g_3^2v^2$, which implies
$v\approx 175/\sqrt{2}$ = 123 GeV. Also, in order to simplify the otherwise cumbersome calculations and to avoid proliferation of unnecessary parameters at this stage of the analysis, we propose to start with the following simple matrix 

\begin{equation} \label{SU}
M^\prime_u=hv\left(\begin{array}{cccc}
0 & 0 & 0 & 1\\
0 & 0 & 0 & 1\\
1 & 1 & h_{32}^u/h & \delta^{-1} \\
1 & 1 &  1  & \delta^{-1} \\
\end{array}\right),
\end{equation}
where $\delta = v/V$ is a perturbation expansion parameter and $h$ is a parameter that can take any value of order one.

Neglecting terms of order $\delta^5$ and higher, the four eigenvalues of $M_u^{\prime\dagger}M_u^\prime$ are: one zero eigenvalue related to the eigenstate $(u^1-u^2)/\sqrt{2}$ (notice the maximal mixing present); one see-saw eigenvalue $4h^2V^2\delta^4 = 4h^2v^2\delta^2\approx m^2_c$ associated to the charm quark mass, and two tree-level values that we identify with the masses of the top quark $t$ and the heavy quark $U$ given, respectively, by 

\[ \frac{h^2V^2\delta^2}{2}[e_-^2+\delta^2e_+^2(4-e_-^2)/4]
\approx \frac{v^2}{2}(h-h_{32}^u)^2\approx m_t^2; \]
\[h^2V^2[2+\delta^2(6+e_+/2)+\delta^4(4e_+^2-e_+^2e_-^2-32)/8] \approx m_U^2, \]
where $e_{\pm}=(1 \pm h_{32}^u/h)$. 

So, in the up quark sector the heavy quark gets a large mass of order $V$ (the 3-3-1 scale), the top quark gets a mass at the electroweak scale [times a difference of Yukawa couplings that in the
general case of the matrix (\ref{matrixU}) is $(h_2^U-h_{32}^u)$], the charm quark gets a see-saw mass, and the first family up quark $u$ remains massless at the tree-level. From the former expressions, and using for $m_t\approx 175$ GeV~\cite{pdb}, we get $|h_2^U-h_{32}^u|\sim 2$ and $m_c\approx 2hv^2/V$, which implies $V\approx hm_t^2/m_c\approx 19.4 h$ TeV, fixing in this way an upper limit for the 3-3-1 mass scale.

\begin{figure}
 \includegraphics{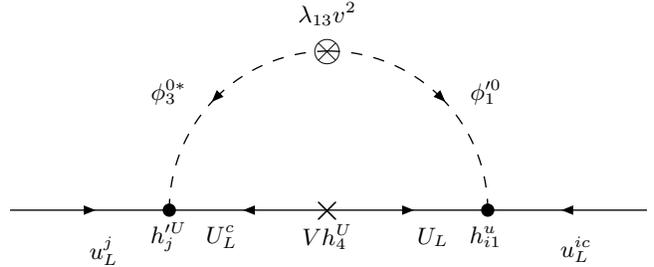}
\caption{One loop diagram contributing to the radiative 
 generation of the up quark mass.}
\label{fig:1}
\end{figure}

The consistency of this model requires to find a mechanism able to produce a mass for the up quark $u$ in the first family. A detailed study of the Lagrangian in Eq.~(\ref{mup}) and the discrete symmetry used, allows us to draw the radiative diagram in Fig.~\ref{fig:1}, which is the only diagram available to produce one-loop radiative corrections in the quark subspace $(u^1,u^2)$. The mixing in the Higgs sector in the diagram comes from a term in the scalar potential of the form $\lambda_{13}(\phi_1^*\phi_1)(\phi_3^*\phi_3)$, which turns on the radiative corrections.

In the analysis we must be careful because, in order to have a contribution different from zero, we must avoid maximal mixing in the first two weak interaction states, otherwise a submatrix of the democratic type arises. This is simply done by taking $h^u_{11} = 1-k$ and $h^{\prime U}_1 = 1+k$ in the matrix (\ref{matrixU}), where $k$ must be a small parameter in order to guarantee the see-saw character of the matrix for the up quarks.

When we evaluate the contribution coming from the diagram in Fig.~\ref{fig:1} we get a finite value given by 

\begin{eqnarray}\label{diag1} \nonumber
\Delta_{ji}&=&N_{ji}[M^2m_1^2\ln (M^2/m_1^2) - M^2m_3^2\ln (M^2/m_3^2)\\
&+&m^2_3m_1^2\ln (m^2_1/m_3^2)],
\end{eqnarray}
where
\begin{equation}
N_{ji}=h_j^{\prime
U}h_{i1}^u\lambda_{13}\frac{v_1v_3M}{16\pi^2(m_3^2-m_1^2)
(M^2-m_1^2)(M^2-m_3^2)}.
\end{equation} 
$M=h_4^UV$ is the mass vertex of the heavy exotic up quark, and $m_1$ and $m_3$ are the masses of $\phi_1^{\prime 0}$
and $\phi_3^0$, respectively. To estimate the contribution given by
this diagram we assume the validity of the ``extended survival
hypothesis" \cite{esh} which, in our case, means $m_1\approx
m_3\approx v<<V\approx M$, which in turn implies a mass value
\[ m_u \approx \,\lambda_{13}v\delta\ln (V/v)/8\pi^2\approx
0.85 \lambda_{13}\, \mbox{MeV},\] 
that for $\lambda_{13}\sim 2$ produces $m_u\approx 1.7$ MeV, which is of the correct order of magnitude~\cite{pdb} (a result independent of the value of $k$ in the first approximation).

Notice that for $k \neq 0$, the state related to the $u$ quark looses its maximal mixing, becoming
now $\{-(h - h_{32}) u^1 + [h- h_{32}(1 -k)] u^2 + k u^3\}/N$, with $N$
being a normalization factor. A value for $k$ can be estimated using the Cabbibo angle.

\subsection{The down quark sector}
The most general Yukawa terms for the down quark sector, 
using the four Higgs scalars introduced in Eq.~(\ref{higgsses}), are 

\begin{eqnarray}\label{mdown} \nonumber
{\cal L}^d_Y &=& \sum_{\alpha=1,2,4}
\sum_{i}Q^i_L\phi_\alpha^*C(\sum_ah^d_{ia\alpha}d_L^{ac}
+\sum_jh^D_{ij\alpha}D_L^{jc})\\
&+& Q_L^3\phi_3C(\sum_ih^D_iD_L^{ic}+\sum_{a}h_a^dd_L^{ac})+h.c..
\end{eqnarray}

In the basis $(d^1,d^2,d^3,D^1,D^2)$ and using the discrete
symmetry $Z_2$, this expression produces the following
tree-level down quark mass matrix 
\begin{equation}\label{MD} M_d=\left(\begin{array}{ccccc}
0 & 0 & 0 & h_{11}^{d}v_1 & h_{21}^{d}v_1 \\
0 & 0 & 0 & h_{12}^{d}v_1 & h_{22}^{d}v_1 \\
0 & 0 & 0 & h_{13}^{d}v_1 & h_{23}^{d}v_1 \\
h_{11}^{D}v_2 & h_{21}^{D}v_2 & h_{1}^{D}v_3 & h_{114}^DV & h_{214}^DV \\
h_{12}^{D}v_2 & h_{22}^{D}v_2 & h_{2}^{D}v_3 & h_{124}^DV & h_{224}^DV \\
\end{array}\right),
\end{equation}
where we have used
$h^{D(d)}_{ia\alpha}v_\alpha=h^{D(d)}_{ia}v_\alpha$. 

The matrix $M_d$ is again a see-saw type mass matrix, with at least one eigenvalue equal to zero, which is plenty of physical possibilities, depending upon the particular values assigned to the Yukawa couplings. For example, if all the Yukawa couplings are different from each other, then the matrix $M_d^\dagger M_d$ has rank one with a zero eigenvalue related to the eingenvector
$[(h_{22}^D h_{1}^D-h_{2}^D h_{21}^D), (h_{11}^D h_{124}^D -
h_{12}^D h_{114}^D), (h_{21}^D h_{12}^D -h_{11}^D h_{22}^D),0,0]$,
that we may identify with the down quark $d$ in the first family (which in any case remains massless at the tree-level). For this case the general analysis shows that we have two see-saw eigenvalues associated with the bottom $b$ and strange $s$ quarks, the first one enhanced by sum of Yukawa couplings and the second one suppressed by differences.

In the particular case when all the Yukawa couplings are equal to one but
$h_{114}^D=h_{224}^D\equiv H^D\neq 1$, the null space of $M_d^\dagger
M_d$ has rank two, with the eigenvectors associated with the zero
eigenvalues given by $[-2,1,1,0,0]/\sqrt{6}$ and
$[0,-1,1,0,0]/\sqrt{2}$, which implies only one see-saw
eigenvalue associated with the bottom quark $b$ with a mass value 
$m_b\approx 6v\delta/(1+H^D)\approx 3m_c/[h(1+H^D)]$, and with masses
for the two heavy states of the order of $V(1\pm H^D)$.

For the first case analyzed in the previous paragraph, the chiral
symmetry remaining at tree-level is $SU(2)_f$ (quarks $u$ and $d$
massless), and for the second case the chiral symmetry is
$SU(3)_f$ (quarks $u$, $d$ and $s$ are massless). In both cases
the chiral symmetry is going to be broken by the radiative corrections.

In any way, a realistic analysis of the down sector requires to
have in mind the mixing matrix of the up quark sector
and the fact that the CKM mixing matrix is almost diagonal and unitary. Aiming to this and in order to avoid again a proliferation of parameters,
let us analyze the particular case given by the following
left-right symmetric (hermitian) down quark mass matrix 

\begin{equation} \label{matrixD}
M^\prime_d=h^\prime v\left(\begin{array}{ccccc}
0 & 0 & 0 & 1 & 1 \\
0 & 0 & 0 & 1 & 1 \\
0 & 0 & 0 & f & g \\
1 & 1 & f & H^D\delta^{-1} & \delta^{-1} \\
1 & 1 & g & \delta^{-1} & H^D\delta^{-1} \\
\end{array}\right),
\end{equation}
where $f$ and $g$ are parameters of order one. This is the
most general hermitian mass matrix with only one zero eigenvalue related with the state $(d^1-d^2)/\sqrt{2}$ (again maximal mixing, as required in
order to end up with an almost diagonal and unitary CKM mixing matrix).

The two see-saw exact eigenvalues of $M^\prime_d$ are 
\begin{eqnarray}
& - & h^\prime\,v\, \frac{\delta}{4}
\left\{\left[\frac{(f-g)^2}{H^D - 1} + \frac{8
+ (f+g)^2}{1+ H^D}\right]\right. \nonumber \\
& \pm & \left.\sqrt{\left[\frac{(f-g)^2}{H^D - 1} + \frac{8 +
(f+g)^2}{1+ H^D}\right]^2 - \frac{8 (f-g)^2}{1-(H^D)^2}}\right\}.
\end{eqnarray}

Moreover, notice that for the particular case $g = -f$ (which implies some Yukawa couplings to become complex), the five eigenvalues of the hermitian matrix above get the following simple exact analytical expressions

\begin{eqnarray}\nonumber
\frac{h^{\prime}\,\delta^{-1}\,v}{2} & &\left[0, H^D_+(1\pm \sqrt{1 +
16 \delta^2/(H^D_+)^2}),\right. \\
& & \left. H^D_-(1\pm \sqrt{1 + 8 f^2\delta^2/(H^D_-)^2})\right],
\end{eqnarray}
where $H^D_{\pm} = 1 \pm H^D$. The two see-saw values are thus
$4\,\delta/H^D_+$ and $ 2\, \delta\,f^2/H^D_-$; which imply $f^2h^\prime/h\approx m_bH^D_-/m_c$ and $2h^\prime/h\approx H^D_+m_s/m_c$, that can be seen as either a mild hierarchy between $h$ and $h^\prime$, or implying a detailed tuning of some of the parameters of order one. The mass of the two heavy states is proportional to $h^\prime VH^D_\pm$.

\begin{figure}
 \includegraphics{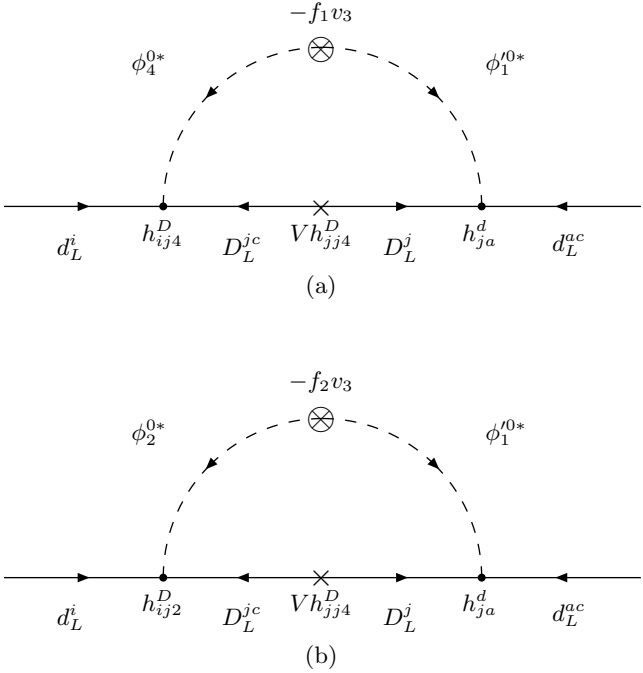}
\caption{One loop diagrams contributing to the radiative 
 generation of the down quark mass.}
\label{fig:2}
\end{figure}

Again, radiative diagrams producing a nonzero mass for the down quark $d$ in the first family must be found. For this purpose we have the four diagrams depicted in Fig.~\ref{fig:2} (two for $D^1$ and other two for $D^2$ in the heavy quark propagator). The mixing in the Higgs sector comes from terms in the scalar potential of the form $(f_1\phi_1\phi_3\phi_4+f_2\phi_1\phi_2\phi_3 + h.c.)$. Now the algebra  shows that 
\begin{equation}
m_d\approx 2(f_1+f_2)\delta \ln (V/v)/8\pi^2, 
\end{equation}
which for $f_1=f_2\approx v$ implies $m_d\approx 2m_u$ without introducing a new mass scale in the model.

\subsection{The lepton sector}
Following the spirit of the analysis in the quark sector, and in order to avoid hierarchies in the Yukawa couplings, we introduce the discrete $Z_2$ symmetry in Eq.~(\ref{z2}) in order to avoid terms proportional to 
\[h_{ab}^eL_{aL}\phi_3Ce_{bL} + h_{ab}^{\prime e}\epsilon_{\alpha\beta\gamma}L_{aL}^\alpha L_{bL}^\beta\phi_3^\gamma + h.c.\]
in the Yukawa Lagrangean. Then, in order to generate masses for the charged leptons we must include either leptoquark Higgs field triplets if we want to use the radiative mechanism, or exotic leptons if we want to use the see-saw mechanism. For example, in Ref.~\cite{mpp2} a singlet exotic charged lepton is introduced in the Pleitez-Frampton model \cite{pf} in order to implement the see-saw mechanism in the lepton sector. This analysis however is outside the scope of the study presented here.

In a similar way, masses for the neutrinos can be generated by introducing either new scalar fields, or new neutral exotic Weyl fermions. For example, in the context of the model studied here, a Majorana mass for the neutrinos can be generated by using scalars belonging to irrep $\{6\}$ of $SU(3)_L$. These scalars can be written as a $3\times 3$ symmetric tensor 
\begin{equation}\label{sextuplet}
\chi_{\alpha\beta} = \left(\begin{array}{ccc}
\chi_{11}^{4/3+X} & \chi_{12}^{1/3+X} & \chi_{13}^{1/3+X} \\
                  & \chi_{22}^{-2/3+X}& \chi_{23}^{-2/3+X} \\
                  &                   & \chi_{33}^{-2/3+X} \\ \end{array}\right)\sim(1,6,X),
\end{equation}
where the upper symbol stands for the electric charge. Clearly, a VEV of the form $\langle\chi_{33}^0(1,6,2/3)\rangle\sim M$ produces a Majorana mass term of the form $M\nu_{lL}^{0c}\nu_{l^\prime L}^{0c}$, 
a VEV of the form $\langle\chi_{22}^0(1,6,2/3)\rangle\sim w$ produces a Majorana mass term of the form $w\nu_{lL}^0\nu_{l^\prime L}^0$, and a VEV of the form $\langle\chi_{23}^0(1,6,2/3)\rangle\sim m$ produces a Dirac mass term for the neutrinos. This issue is studied for example in the several papers in Refs.~\cite{kita1,kita2}, where $SU(3)_L$ scalar singlets, triplets and sextuplets are used in order to provide the model with a realistic neutrino mass spectrum.

\section{Gauge coupling Unification}
\label{sec:3}
In a field theory, the coupling constants are defined as effective values which are energy scale dependent according to the renormalization group equation. In the modified minimal substraction scheme~\cite{mss}, which we adopt in what follows, the one loop renormalization group equation (RGE) for $\alpha = g^2/4\pi$ reads
\begin{equation}\label{rge}
\mu\frac{d\;\alpha}{d\;\mu}\simeq -b\alpha^2,
\end{equation}
where $\mu$ is the energy at which the coupling constant $\alpha$ is evaluated. The constant value $b$, called the beta function, is completely determined by the particle content of the model by 
\[2\pi b=\frac{11}{6}C(\mbox{vectors})-\frac{2}{6}C(\mbox{fermions})-\frac{1}{6}C(\mbox{scalars}),\]
where $C(\dots )$ is the group theoretical index of the representation inside the parentheses (we are assuming Weyl fermions and complex scalar fields~\cite{slansky}).

For the energy interval $m_Z<\mu<M_G$, the one loop solutions to the RGE (\ref{rge}) for the three SM gauge coupling constants are
\begin{equation}\label{srge}
\alpha^{-1}_i(m_Z)=\frac{\alpha_i^{-1}(M_G)}{c_i}-b_i(F,H)\ln\left(\frac{M_G}{m_Z}\right), 
\end{equation}
where $i=Y,2,c$ refers to the coupling constants of $U(1)_Y$, $SU(2)_L$ and $SU(3)_c$, respectively, with the beta functions given by 
\begin{equation}
2\pi\left(\begin{array}{c} b_Y \\b_2 \\b_c \end{array}\right)=
\left(\begin{array}{c} 0 \\ \frac{22}{3} \\ 11 \end{array}\right)
-\left(\begin{array}{c} \frac{20}{9} \\ \frac{4}{3} \\ \frac{4}{3} \end{array}\right)F
-\left(\begin{array}{c} \frac{1}{6} \\ \frac{1}{6} \\ 0 \end{array}\right)H, 
\end{equation}
where $F$ is the number of families contributing to the beta functions and $H$ is the number of low energy $SU(2)_L$ scalar field doublets ($H=1$ for the SM). In Eq.~(\ref{srge}) the constants $c_i$ are group theoretical factors which depend upon the embedding of the SM factors into a covering group, and warrant the same normalization for the covering group $G$ and for the three group factors in the SM. For example, if the covering group is $SU(5)$, then $(c_Y,c_2,c_c)=(3/5,1,1)$, but they are different for other covering groups (see for example the Table in Ref.~\cite{lpz}).

The three running coupling constants in $\alpha_i$, may or may not converge into a single energy GUT scale $M_G$; if they do, then $\alpha_i(M_G)=\alpha$ is a constant independent of the index $i$. Now, for a given embedding into a fixed covering group, the $c_i$ values are fix, and if we use for $F=3$ (an experimental fact) and $H=1$ as in the SM, then  Eq.~(\ref{srge}) constitute a set of three equations with two unknowns, $\alpha$ and $M_G$, which may or may not have a consistent solution (more equations than unknowns). 

The inputs to be used in Eq.~(\ref{srge}) for $\alpha^{-1}_i(m_Z)$ are calculated from the experimental results~\cite{pdb} 
\begin{eqnarray}\nonumber
\alpha_{em}^{-1}(m_Z)&=& \alpha_Y^{-1}(m_Z)+\alpha_2^{-1}(m_Z)\\ \nonumber 
&=& 127.918\pm 0.018\\ \nonumber
\sin^2\theta_W(m_Z)&=&1-\alpha_Y^{-1}(m_Z)\alpha_{em}(m_Z) \\ \nonumber 
&=&0.23120\pm 0.00015\\ \nonumber
\alpha_c(m_Z)&=& 0.1213\pm 0.0018,
\end{eqnarray}
which imply $\alpha^{-1}_Y(m_Z)=98.343\pm 0.036, \;\alpha_2^{-1}(m_Z)=29.575 \pm 0.054$, and $\alpha_c^{-1}(m_Z)= 8.244 \pm 0.122$.

It is a well known fact that the model based on the nonsupersymmetric $SU(5)$ group of Georgi and Glashow~\cite{gg} lacks of gauge coupling unification because $M_G$ is not unique in the range $10^{14}$ GeV $\leq M_G \leq 10^{16}$ GeV, predicting for the proton lifetime $\tau_p$ a value between $2.5\times 10^{28}$ years and $1.6\times 10^{30}$ years, which by the way is ruled out by experimental measurements~\cite{amaldi}. If we introduce one more free parameter in the solutions to the RGE, as for example letting $H$ to become a free integer number, then we have now three unknowns with three equations that always have mathematical solution (not necessarily with physical meaning). Doing that in Eqs.~(\ref{srge}) we find that for $H=7$ (seven Higgs doublets) we get the unique solution $M_G=10^{13}$ GeV $>>m_Z$ which, altough a physical solution, is ruled out by the proton lifetime. So, if we still want unification, new physics at an intermediate mass scale $M_V$ such that $m_Z<M_V<M_G$ must exists, being supersymmetry (SUSY) a popular candidate for that purpose~\cite{amaldi}. 

The question now is if the 3-3-1 model under consideration in this paper, introduces an intermediate mass scale $M_V$ such that it achieves proper gauge coupling unification, being an alternative for SUSY. To answer this question using $SU(6)$ as the covering group as presented in 
Sect.~\ref{sec:1}, we must solve the following set of seven equations:
\begin{eqnarray}\label{srge331}\nonumber
\alpha^{-1}_i(m_Z)&=&\frac{\alpha_i^{-1}(M_V)}{c_i}-b_i(F,H)\ln\left(\frac{M_V}{m_Z}\right)\\ \nonumber
\alpha^{-1}_j(M_V)&=&\frac{\alpha^{-1}}{c_j^\prime}-b_j^\prime\ln\left(\frac{M_G}{M_V}\right)\\ 
\alpha_Y^{-1}(M_V)&=&\alpha_1^{-1}(M_V)+\alpha_3^{-1}(M_V)/3,
\end{eqnarray}
where the last equation is just the matching conditions in Eq.~(\ref{mc}), and $i=c,2,Y$ and $j=c,3,1$ for the SM and the 3-3-1 model, respectively. The constants $c_i$ are $(c_Y,c_2,c_3)=(3/5,1,1)$ as before, and $(c_1^\prime,c_3^\prime,c_c^\prime)=(3/4,1,1)$, with the value $c_1^\prime =3/4$ calculated from the electroweak mixing angle in Eq.~(\ref{sinteta}). $b_j^\prime$ stand for the beta functions for the 3-3-1 model under study here.

Eqs.~(\ref{srge331}) constitute a set of seven equations with seven unknowns $\alpha, \; \alpha_j(M_V), \; M_V, \; M_G$ and $\alpha_Y(M_V) \;\; [\alpha_2(M_V)=\alpha_3(M_V)$ according to the matching conditions]. There is always mathematical solution to this set of equations, but we want only physical solutions, that is solutions such that $m_Z<M_V<M_G$.

The new beta functions, calculated with the particle content introduced in Sect.~\ref{sec:1}, are 
\begin{equation}\label{beta331}
2\pi\left(\begin{array}{c} b_1^\prime \\b_3^\prime \\b_c^\prime \end{array}\right)=
\left(\begin{array}{c} 0-8-7/9 \\ 11-4-4/6 \\ 11-6-0 \end{array}\right)
=\left(\begin{array}{c} -79/9 \\ 19/3 \\ 5 \end{array}\right),
\end{equation}
where in the middle term we have separated the contributions coming from the gauge bosons, the fermion fields and the scalar fields in that order. When we introduce these values in Eq.~(\ref{srge331}), we do not obtain a physical solution in the sense that we get $m_Z<M_G<M_V$.

Of course, if there are more particles at the 3-3-1 mass scale then the beta functions given in Eqs.~(\ref{beta331}) are not the full story. In particular we know from Sect.~\ref{sec:2} that at least new Higgs scalars are needed in order to generate a consistent lepton mass spectrum, so let us allow the presence in the model of the following Higgs scalar multiplets at the 3-3-1 mass scale: $N_X^{(1)} \; SU(3)_L$ singlets (with $U(1)_X$ hypercharge equal to $X$), $N_X^{(3)}$ triplets (color singlets), $\tilde{N}_X^{(3)}$ leptoquark triplets (color triplets) and $N_X^{(6)}$ sextuplets (color singlets). These new particles will contribute to the beta functions $b^\prime_j$ in the following way:

\begin{equation}\label{betap331}
2\pi\left(\begin{array}{c} b_1^\prime \\ b_3^\prime \\b_c^\prime \end{array}\right)=
\left(\begin{array}{c} 
-79/9 -\sum_XX^2 f(N_X^{(6)}, \tilde{N}_X^{(3)}, N_X^{(3)},N_X^{(0)})\\  
19/3-{1 \over 6}\sum_X(N_X^{(3)}+3\tilde{N}_X^{(3)}+5N_X^{(6)}) \\ 
5-\sum_X\tilde{N}_X^{(3)}/2 \end{array}\right), 
\end{equation}
where $f(N_X^{(6)}, \tilde{N}_X^{(3)}, N_X^{(3)},N_X^{(0)})=  
(2N_X^{(6)}+3\tilde{N}_X^{(3)}+N_X^{(3)}+N_X^{(0)}/3)$; 
with these new $SU(3)_L$ multiplets contributing or not to the beta functions $b_i$ of the SM factor groups, in agreement with the extended survival hypothesis~\cite{esh} (for example, a sextuplet with a VEV $\langle\chi_{23}(1,6,2/3)\rangle \sim v$ contributes as an $SU(2)_L$ doublet in $b_Y$ and $b_2$, etc.).

\begin{figure*}
 \includegraphics{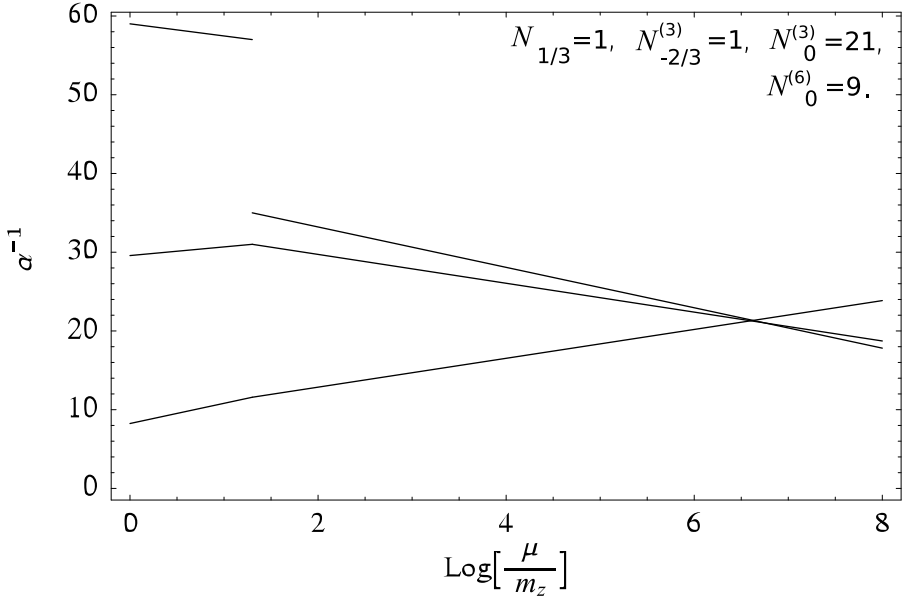}
\caption{Solutions to the RGE for the 3-3-1 model. For the meaning of $N^{(r)}_X$ see the main text.}
\label{fig:3}
\end{figure*}

The calculation shows that for the following set of extra scalar Higgs fields that do not develop VEV: $N_X^{(0)}=0$, $N^{(3)}_{1/3}=1$, $N^{(3)}_{-2/3}=1$, $\tilde{N}^{(3)}_X=0$ $N^{(3)}_0=21$ and $N_0^{(6)}=9$, the set of equations in (\ref{betap331}) has the physical solution 
\begin{equation}\label{scales}
M_V\approx 1.9 \mbox{TeV}  < M_G\approx 5\times 10^{8} \mbox{GeV}, 
\end{equation}
which provides with a convenient 3-3-1 mass scale, and a low unification GUT mass scale, as it is shown in Fig.~\ref{fig:3}.

But, is this low GUT scale in conflict with the bounds on proton decay?. The answer is not, because due to the $Z_2$ symmetry our unifying group is $SU(6)\times Z_2$. Then we must assign to each irrep of $SU(6)$ in Eq.~(\ref{rred}) a given $Z_2$ charge in accord with the $Z_2$ value assigned to the 3-3-1 states in Eq.~(\ref{z2}). For example, if we assign to one of the eight $\{6^*\}=\{D^c, -N_E^0,E^-,N_E^{0c}\}_L$ states in (\ref{rred}) a $Z_2$ value equal to 1, then we can perfectly identify $D_L^c$ with one of the ordinary down quarks $(d^c,s^c,b^c)_L$, but then $(-N_E^0, E^-, N_E^{0c})_L$ can not correspond to $(-\nu_l^0, l^-,\nu_l^{0c})_L$ because all of them have a $Z_2$ charge equal to zero; and the same for the other way around. As a consequence, the down quark $d_L^c$ can not live together with $(\nu_e,e^-)_L$ in the same $SU(6)\times Z_2$ irrep, and the proton can not decay into light states belonging to the weak basis. The decay can of course occur via the mixing of ordinary 3-3-1 states with the extra new states in $SU(6)$, but such a mixing is of the order of $(M_V/M_G)^2$ which is a very small value. Of course, this argument is valid as far as we can find a mechanism able to produce GUT scale masses for all the extra states, but such analysis is outside the present work.

\section{Constraints on the parameters}
\label{sec:4}
In this section we are going to set bounds on the mass of the new neutral gauge boson $Z_2^\mu$, and its mixing angle $\theta$ with the ordinary neutral gauge boson, using the partial decay width for $Z_1^\mu$. We also are going to set constraints coming from possible FCNC effects, and to analyze the violation of unitarity of the Cabbibo-Kobayashi-Maskawa mixing matrix $V^0_{CKM}$.

\subsection{Bounds on $M_{Z_2}$ and $\theta$}
Let us notice to start that, after the identification of the mass eigenstates, we can properly
bound $\sin\theta$ and $M_{Z_2}$ by using parameters measured at the $Z$
pole from CERN $e^+e^-$ collider (LEP), SLAC Linear Collider (SLC), and
atomic parity violation constraints which are given in Table~\ref{tab:3}.

The expression for the partial decay width for $Z^{\mu}_1\rightarrow
f\bar{f}$ is
 
\begin{eqnarray}\nonumber
\Gamma(Z^{\mu}_1\rightarrow f\bar{f})&=&\frac{N_C G_F
M_{Z_1}^3}{6\pi\sqrt{2}}\rho \Big\{\frac{3\beta-\beta^3}{2}
[g(f)_{1V}]^2 \\ \label{ancho}
& + & \beta^3[g(f)_{1A}]^2 \Big\}(1+\delta_f)R_{EW}R_{QCD}, \quad
\end{eqnarray}
\noindent 
where $f$ is an ordinary SM fermion, $Z^\mu_1$ is the physical gauge boson
observed at LEP, $N_C=1$ for leptons while for quarks
$N_C=3(1+\alpha_s/\pi + 1.405\alpha_s^2/\pi^2 - 12.77\alpha_s^3/\pi^3)$,
where the 3 is due to color and the factor in parentheses represents the
universal part of the QCD corrections for massless quarks 
(for fermion mass effects and further QCD corrections which are 
different for vector and axial-vector partial widths, see 
Ref.~\cite{kuhn}); $R_{EW}$ are the electroweak corrections which include 
the leading order QED corrections given by $R_{QED}=1+3\alpha/(4\pi)$. 
$R_{QCD}$ are further QCD corrections (for a comprehensive review see 
Ref.~\cite{leike} and references therein), and $\beta=\sqrt{1-4 m_f^2/
M_{Z_1}^2}$ is a kinematic factor which can be taken equal to $1$ for all 
the SM fermions except for the bottom quark. 
The factor $\delta_f$ contains the one loop vertex
contribution which is negligible for all fermion fields except for the 
bottom quark, for which the contribution coming from the top quark, at the 
one loop vertex radiative correction, is parameterized as $\delta_b\approx 
10^{-2} [-m_t^2/(2 M_{Z_1}^2)+1/5]$~\cite{pich}. The $\rho$ parameter 
can be expanded as $\rho = 1+\delta\rho_0 + \delta\rho_V$ where the 
oblique correction $\delta\rho_0$ is given by
$\delta\rho_0\approx 3G_F m_t^2/(8\pi^2\sqrt{2})$, and $\delta\rho_V$ is 
the tree level contribution due to the $(Z_{\mu} - Z'_{\mu})$ mixing which 
can be parameterized as $\delta\rho_V\approx
(M_{Z_2}^2/M_{Z_1}^2-1)\sin^2\theta$. Finally, $g(f)_{1V}$ and $g(f)_{1A}$
are the coupling constants of the physical $Z_1^\mu$ field with ordinary
fermions which, for this model, are listed in Table~\ref{tab:1}.

In what follows we are going to use the experimental values~\cite{pdb}:
$M_{Z_1}=91.188$ GeV, $m_t=174.3$ GeV, $\alpha_s(m_Z)=0.1192$,
$\alpha(m_Z)^{-1}=127.938$, and $\sin\theta^2_W=0.2333$. These
values are introduced using the definitions $R_\eta\equiv
\Gamma_Z(\eta\eta)/\Gamma_Z(hadrons)$ for $\eta=e,\mu,\tau,b,c,s,u,d$.

As a first result, notice from Table~\ref{tab:1} that this model predicts 
$R_e=R_\mu=R_\tau$, in agreement with the experimental results in 
Table~\ref{tab:3}, independent of any flavor mixing at the tree-level.

The effective weak charge in atomic parity violation, $Q_W$, can be 
expressed as a function of the number of protons $(Z)$ and the number of 
neutrons $(N)$ in the atomic nucleus in the form 

\begin{equation}
Q_W=-2\left[(2Z+N)c_{1u}+(Z+2N)c_{1d}\right], 
\end{equation}
\noindent
where $c_{1q}=2g(e)_{1A}g(q)_{1V}$. The theoretical value for $Q_W$ for 
the cesium atom is given by~\cite{ginges} $Q_W(^{133}_{55}Cs)=-73.19\pm 0.13 + \Delta Q_W$, where the contribution of new physics is included in $\Delta Q_W$ which can be written as~\cite{durkin}

\begin{equation}\label{DQ} 
\Delta 
Q_W=\left[\left(1+4\frac{S^4_W}{1-2S^2_W}\right)Z-N\right]\delta\rho_V
+\Delta Q^\prime_W.
\end{equation}

The term $\Delta Q^\prime_W$ is model dependent and it can be obtained for
our model by using $g(e)_{iA}$ and $g(q)_{iV}$, $i=1,2$, from Tables~\ref{tab:1} and \ref{tab:2}. The value we obtain is

\begin{equation}
\Delta Q_W^\prime=(3.75 Z + 2.56 N) \sin\theta + (1.22 Z + 0.41 N)
\frac{M^2_{Z_1}}{M^2_{Z_2}}\; .
\end{equation}

The discrepancy between the SM and the experimental data for $\Delta Q_W$ 
is given by~\cite{ginges}

\begin{equation}
\Delta Q_W=Q^{exp}_W-Q^{SM}_W=0.45\pm 0.48,
\end{equation}
which is $1.1\; \sigma$ away from the SM predictions.

\begin{table}
\caption{Experimental data and SM values for some parameters 
related with neutral currents.}
\label{tab:3}
\begin{tabular}{lcl}
\hline\noalign{\smallskip}
& Experimental results & SM \\ 
\noalign{\smallskip}\hline\noalign{\smallskip}
$\Gamma_Z$(GeV)  & $2.4952 \pm 0.0023$  &  $2.4966 \pm 0.0016$  \\   
$\Gamma(had)$ (GeV)  & $1.7444 \pm 0.0020$ & $1.7429 \pm 0.0015$ \\ 
$\Gamma(l^+l^-)$ (MeV) & $83.984\pm 0.086$ & $84.019 \pm 0.027$ \\
$R_e$ & $20.804\pm 0.050$ & $20.744\pm 0.018$ \\ 
$R_\mu$ & $20.785\pm 0.033$ & $20.744\pm 0.018$ \\ 
$R_\tau$ & $20.764\pm 0.045$ & $20.790\pm 0.018$ \\ 
$R_b$ & $0.21664\pm 0.00068$ & $0.21569\pm 0.00016$ \\ 
$R_c$ & $0.1729\pm 0.0032$ & $0.17230\pm 0.00007$ \\ 
$Q_W^{Cs}$ & $-72.74\pm 0.29\pm 0.36$ & $-73.19\pm 0.13$  \\
$M_{Z_{1}}$(GeV) & $ 91.1872 \pm 0.0021 $ & $ 91.1870 \pm 0.0021 $ \\ 
\noalign{\smallskip}\hline
\end{tabular}
\end{table}

\begin{figure*}
 \includegraphics{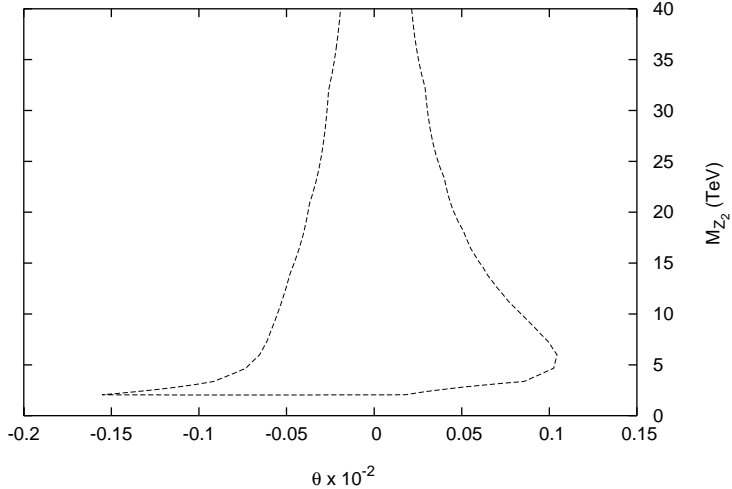}
\caption{Contour plot displaying the allowed region for 
 $\theta$ vs $M_{Z_2}$ at 95\% C.L..}
\label{fig:4}
\end{figure*}

Introducing the expressions for $Z$ pole observable in Eq.~(\ref{ancho}),
with $\Delta Q_W$ in terms of new physics in Eq.~(\ref{DQ}) and using
experimental data from LEP, SLC and atomic parity violation (see 
Table~\ref{tab:3}), we do a $\chi^2$ fit and we find the best allowed region in the $(\theta-M_{Z_2})$ plane at $95\%$ confidence level (C.L.). 
In Fig.~\ref{fig:4} we display this region which gives us the constraints
\begin{equation}\label{graph}
-0.00156\leq\theta\leq 0.00105, \quad \qquad 2.1\; {\mbox TeV} \leq
M_{Z_2}. 
\end{equation}
As we can see, the mass of the new neutral gauge boson is compatible with
the bound obtained in $p\bar{p}$ collisions at the Fermilab Tevatron
~\cite{abe}. From our analysis we can also see that for $\vert \theta \vert
\rightarrow 0$, $M_{Z_2}$ peaks at a finite value larger than $100$~TeV
which still copes with the experimental constraints on the $\rho$
parameter.

\subsection{Bounds from Unitarity Violation of the CKM mixing matrix}
The see-saw mass mixing matrices for quarks presented in 
Eqs.~(\ref{matrixU}) and (\ref{MD}) are not a consequence of the particular discrete $Z_2$ symmetry introduced in Eq.~(\ref{z2}), in the sense that it is a straightforward calculation to show that any $Z_n$ symmetry will reproduce the same quark mass matrices as far as we impose the following constraints:
\begin{itemize}
\item To have a pure see-saw mass matrix in the down quark sector.
\item To have a tree-level mass entry for the top quark in the third family, plus a see-saw matrix for the other two families in the up quark sector.
\item To work with the non-minimal set of four Higgs scalar fields introduced in Eq.~(\ref{higgsses}).
\end{itemize}

As a consequence of the mixing in the quark mass matrices, violation of unitarity of the CKM mixing matrix appears. Notice that for this particular model, $V^0_{CKM}$ is obtained as the upper left $3\times 3$ submatrix of a $4\times 5$ mixing matrix (obtained, in turn, as the product of a $4\times 3$ submatrix taken from the unitary $4\times 4$ diagonalization matrix in the up quark sector with the fourth column suppressed, times a $3\times 5$ submatrix of the unitary $5\times 5$ diagonalization matrix of the down quark sector with the last two rows suppressed, all this as a consequence of having only three active quarks in the charged weak current $J_{W^+}^\mu$). 

The unitarity violation arising in the model must be compatible with the experimental constraints on the CKM mixing parameters, as discussed, for example, in section 11 of Ref.~\cite{pdb}, where uncertainties in the third decimal place of the entries $V^0_{u_id_j}$ (the $i,j$ element of $V^0_{CKM}$), can be taken as possible signals of violation of unitarity.

Now, for the model discussed here, the structure of the quark mass matrices implies a mixing proportional to $\cos\delta$ (with $\delta=v/V$, as before) for the known quarks of each sector, which, when combined in the $V^0_{u_i d_j}$ entries, gives a mixing of the form $\cos^2\delta=1-\sin^2\delta\approx 1-\delta^2$, being $\delta^2$ proportional to the violation of unitarity in the model. Taking for $V\approx M_{Z_2}\approx 2.1$ TeV (the lower bound in Eq.~(\ref{graph})), we obtain $\delta^2\approx 3.4 \times 10^{-3}$, which is in the limit of the allowed unitarity violation of $V^0_{CKM}$~\cite{pdb}.

But the former is not the full story, because violation of unitarity of $V^0_{CKM}$ automatically induces FCNC. Unfortunately, violation of unitarity of $V^0_{CKM}$ is not the best place to look for FCNC processes, because almost all the phenomenology of the CKM mixing matrix is done under the assumption of unitarity, which is not the case in the model presented here.

\subsection{FCNC Processes}
In a model like this, with four scalar triplets and mixing of ordinary with exotic fermion fields, we should worry about possible FCNC effects.

First, notice that due to our $Z_2$ symmetry, FCNC do not occur at tree-level in the Lagrangian, because each flavor couples only to a single multiplet. But FCNC effects can occur in $J_{\mu ,L}(Z)$ and $J_{\mu ,L}(Z^\prime)$ in Eqs.~(\ref{jotaz}) and (\ref{jzprima}), respectively, due again to the mixing of ordinary and heavy exotic fermion fields (notice from Eq.~(\ref{jotaz}) that $J_{\mu ,L}(Z)$ only includes as active quarks the three ordinary up- and down-type quarks).

The best place to study the suppression of $d\leftrightarrow s$ currents is in the $(K^0_L-K^0_S)$ mass difference, which may get contributions from the exchange of $Z_1$ and $Z_2$ between $d\leftrightarrow s$ currents. The contribution from $Z_1$ is proportional to $|V^{0\dagger}_{us}V^0_{ud}|^2\approx |V^{0\dagger}_{us}V^0_{ud}|^2_{SM}+4\delta^4$ (where $|V^{0\dagger}_{us}V^0_{ud}|^2_{SM}$ refers to the SM contribution which is in agreement with the experimental data). Then, the mixing of light and heavy quarks implies extra FCNC effects proportional to $4\delta^4$, which, for $V\approx 2.1$ TeV as before, implies a contribution to new FCNC effects proportional to $1.2\times 10^{-5}$. This value should be compared with the experimental bound $m(K_L)-m(K_S)\approx 3.48\pm 0.006\times 10^{-12}$ MeV \cite{pdb}. Then, for $V\approx 2.1$ TeV we have that $4\delta^4\leq 0.006/3.48$, which means that there is room in the experimental uncertainties to include the new FCNC effects coming from violation of unitarity of the CKM mixing matrix present in the model.

Now, the contributions coming from $Z_2$ alone are safe, because they are not only constrained by the $\delta$ parameter, but also by the mixing angle $-0.00156\leq \theta\leq 0.00105$ as given by Eq.~(\ref{graph}).


\section{Conclusions}
During the last decade several 3-3-1 models for one and three families
have been analyzed in the literature, the most popular one being the
Pleitez-Frampton model~\cite{pf} which certainly is not the only 
possible construction based on this local gauge group. 
Other two different three-family models, more appealing but not so popular
in the literature, are introduced in Refs.~\cite{vl} and \cite{ozer,pgs}. The model in Ref.~\cite{vl}, studied in this paper, contains right-handed neutrinos, while the model in Ref.~\cite{ozer} does not include right-handed neutrinos but it has one extra exotic electron per family. Even more, the analysis presented in Refs.~\cite{pfs,pgs} shows that indeed, there are an infinite number of anomaly-free models based on the 3-3-1
gauge structure, most of them including particles with exotic electric
charges; but the number of models with particles without exotic electric
charges are just a few. For example, other two 3-3-1 models for one family and only with particles of ordinary electric charge, are analyzed in Refs.~\cite{spm}.

In this paper we have carried out a systematic study of the so called 3-3-1 model with right handed neutrinos. In concrete, we have recalculated its charged and neutral currents, embedded the structure into $SU(6)$ as a covering group, looked for unification possibilities, studied the quark mass spectrum, and finally, by using updated precision measurements of the electroweak sector, we have set new limits for the mixing angle between the two heavy electrically neutral gauge weak bosons.

In our analysis we have done a detailed study of the conditions that produce a consistent quark mass spectrum in the context of this model, an analysis only sketched in previous works~\cite{vl}, except for the neutral lepton sector \cite{kita1}. First we have shown that a set of four Higgs scalars is enough to properly break the symmetry producing a consistent mass spectrum in the gauge boson sector. Then, the introduction of an appropriate anomaly-free discrete $Z_2$ symmetry allows us to construct an appealing mass spectrum in the quark sector without hierarchies in the Yukawa couplings. In particular we have carried a program in which: the three exotic quarks get heavy masses at the TeV scale; the top quark gets a tree-level mass at the electroweak scale; then the bottom, charm and strange quarks get see-saw masses, and finally, the first family quarks get radiative masses in such a way that $m_d\approx 2m_u$; the former without introducing strong hierarchies in the Yukawa coupling constants, neither new mass scales in the model.

In addition, we have also embedded the model into the covering group $SU(6)\supset SU(5)$ and studied the conditions for gauge coupling unification at a scale $M_G\approx 5\times 10^8$ GeV. The analysis has shown that a physical $(m_Z<M_V<M_G)$ one loop solution to the RGE can be achieved at the expense of introducing extra Higgs scalars at the intermediate energy scale $M_V$.

The fact that the RGE produces a 3-3-1 mass scale of the same order ($\sim 2$ TeV) than the lower limit obtained in the phenomenological analysis presented in Sect.~\ref{sec:4} [compare Eqs.~(\ref{scales}) and (\ref{graph})] is not accidental neither fortuitous. As a mater of fact, the extra scalar fields contributing to the beta functions in Eq.~(\ref{betap331}), were just introduced for doing this job. A different set of scalar Higgs fields will produce either a different 3-3-1 and GUT mass scales, not unification at all, or either unphysical solutions. Eventhough our analysis may look a little arbitrary, we emphasize that we take the decision of playing only with the most obscure part of any local gauge theory: the Higgs scalar sector.

Finally, we want to stress that, as discussed in the previous section, the lower bound $2.1$ TeV for the mass of the new neutral gauge boson $Z^\mu_2$ is compatible with the constraints coming from violation of unitarity of the CKM mixing matrix and from new contributions to FCNC processes.

\section*{Acknowledgments}
We acknowledge partial financial support from DIME at Universidad Nacional de Colombia-Sede Medell\'\i n, and from CODI at Universidad de Antioquia. W.A. Ponce thanks the {\it ''Laboratorio de f\'\i sica te\'orica de la Universidad de La Plata}" in La Plata, Argentina, where part of this work was done. We also thank C. Garc\'\i a-Canal, Huner Fanchiotti and  Enrico Nardi for illuminating discussions.

\end{document}